# Ge/Si nanowire mesoscopic Josephson junctions

Jie Xiang[1], A. Vidan[2,*], M. Tinkham[2,3], R. M. Westervelt[2,3,†] & Charles M. Lieber[1,2,†]

[1]*Department of Chemistry and Chemical Biology, Harvard University, Cambridge, MA, 02138, USA*

[2]*Division of Engineering and Applied Sciences, Harvard University, Cambridge, MA, 02138, USA*

[3]*Department of Physics, Harvard University, Cambridge, MA, 02138, USA*

[*]*Present address: MIT Lincoln Laboratory, Lexington, Massachusetts, 02420, USA*

[†]*e-mail: westervelt@deas.harvard.edu, cml@cmliris.harvard.edu*

**Abstract**

*The controlled growth of nanowires (NWs) with dimensions comparable to the Fermi wavelengths of the charge carriers allows fundamental investigations of quantum confinement phenomena. Here, we present studies of proximity-induced superconductivity in undoped Ge-Si core-shell NW heterostructures contacted by superconducting leads. By using a top gate electrode to modulate the carrier density in the NW, the critical supercurrent can be tuned from zero to greater than 100 nA. Furthermore, discrete subbands form in the NW due to confinement in the radial direction, which results in step-wise increases in the critical current as a function of gate voltage. Transport measurements on these superconductor-NW-superconductor devices reveal high order (n=25) resonant multiple Andreev reflections, indicating that the NW channel is smooth and the charge transport is highly coherent. The ability to create and control coherent superconducting ordered states in semiconductor-superconductor hybrid nanostructures allows for new opportunities in the study of fundamental low-dimensional superconductivity.*

When two superconductors are weakly connected, for example in a superconductor-normal conductor-superconductor (S-N-S) structure, a dissipationless supercurrent can flow through the junction as a result of the Josephson effect, which originates from the fixed phase differences between the electron wavefunctions in the two superconductors across the normal conductor[1,2]. This type of device is known as the Josephson junction, with the maximum magnitude of the supercurrent flowing across the junction defined as the Josephson critical current, $I_c$. In the clean limit of a conventional metallic junction[1], the product of $I_c$ with normal state resistance, $R_n$, is a constant proportional to the BCS energy $\Delta$: $I_c R_n = \pi\Delta/e$, where $e$ is the fundamental charge of an electron. The $I_c R_n$ product is independent of sample geometry, since the same junction geometry dependent terms cancel each other in $I_c$ and $R_n$ [1,2]. Interestingly, a new mesoscopic regime emerges when the width, $w$, of the normal conductor shrinks to become comparable to the carrier Fermi wavelength, $\lambda_F$, and its normal state conductance becomes quantized in multiples of $2e^2/h$, regardless of the exact length of the constriction due to the effect of one-dimensional (1D) quantum confinement[3]. It has been predicted[4] that the universal product $I_c R_n = \pi\Delta/e$ also holds for short Josephson junctions with discrete transverse modes, where each of the $N$ modes forms an independent Andreev bound level and contributes the same amount to the total supercurrent[4,5], thus $I_c = Ne\Delta/\hbar$, although such a regime has not been reached experimentally[6,7]. In most previous investigations of S-N-S structures, conventional metals have been used to form the junctions. In these junctions, the $w \sim \lambda_F$ regime is difficult to achieve as it would require a stable and controllable junction only a few atoms wide[8]. This limitation can be overcome using semiconductors due to their low carrier density and correspondingly larger Fermi wavelength. Previous experimental studies with Nb-contacted quantum point contacts (QPCs) in a two-dimensional electron gas[6,7] reported steps in $I_c$, although these were two orders of magnitude smaller than $e\Delta/\hbar$ and dependent on the sample geometry due to the large sample size ($L \gg \xi_0$, where $\xi_0$ is the coherence length characteristic of the superconductor). The nanoscale dimensions and tunability of carrier density with applied gate voltage in chemically synthesized semiconductor NWs make them promising platforms to carry out such studies.

Indeed, recent advances in the chemical growth and physical understanding of NWs[9,10] have enabled the study of a broad range of transport phenomena associated with quantum confinement effects and reduced dimensionality, such as single charge tunneling[11-13] and coherent ballistic transport with controlled access of multiple 1D subbands[14], and have enabled high performance field effect transistors[15]. Furthermore, recently the pioneering work by Kouwenhoven and coworkers has demonstrated supercurrents in n-type InAs NWs[16] and has enabled the study of supercurrents coupled with Coulomb blockade phenomena in quantum dots[17], although the relatively large diameters and small mean free paths in these NWs suggest that carrier transport was in the bulk diffusive regime. On the other hand, epitaxial Ge/Si core/shell NW heterostructures[14,15] (Fig. 1a) have yielded a low-dimensional carrier gas with enhanced mobility, low scattering, and reproducible ohmic contacts. These characteristics, along with the ability to tune conduction down to the first few 1D modes in the quantum confined Ge channel make Ge/Si NWs ideal for the study of the interplay between 1D quantum confinement and superconductivity. In this paper we utilise superconducting aluminium as contacts to individual Ge/Si NWs in order to investigate this novel regime (See Fig. 1(c,d)). Measurements on a single S-NW-S device found a tunable critical supercurrent $I_c$ of greater than 100 nA. At finite bias voltage the device exhibited resonant multiple Andreev reflection peaks in differential conductance up to order 25, indicating that the NW channel is smooth and transport is highly coherent[5,18,19]. Finally, we found that $I_c$ exhibits steps at gate voltage values corresponding to the normal state conductance plateaus due to radial carrier confinement.

## Results and Discussion

### Fabrication of Ge/Si NW devices

The Ge/Si core/shell NWs were grown using a two-step chemical vapor deposition process, and have a Ge core diameter of approximately 15 nm and Si shell thickness of 1.5~2 nm[15,20]. As shown in previous studies[14,15], the epitaxial Si shell provides a clean confinement potential for the hole gas in the Ge core based on their ~500 mV valence band differences (Fig. 1b). After growth, NW devices with Al contacts were fabricated by electron beam lithography and thermal evaporation with typical contact separation length, which defines the length $L$ of the Josephson junction, of 100 to 150 nm (Methods). A top-gated structure with high-κ dielectric[15] was used to control the Fermi level and hole carrier density inside the NW (Figs. 1c,d). The contacts were not intentionally annealed, although we believe Fermi level pinning due to interfacial states and diffusion of aluminum during the fabrication process are responsible for the ohmic contacts to the Ge hole gas at low temperature.

### Normal state characteristics

Figure 2a shows the differential conductance, $dI/dV$, as a function of source-drain voltage $V$ in a 150 nm long NW. The measurement was performed at 10 K so that the aluminium leads are non-superconducting. The different curves correspond to different values of the top gate voltage $V_g$. In the linear regime (around $V = 0$), dark bunching regions, where several curves at different $V_g$ overlap, are observed and are spaced vertically by approximately $2e^2/h$, which are consistent with quantized conductance plateaus for individual spin degenerate 1D subbands[3], since $V_g$ has little effect on $dI/dV$ in the plateau regions. The irregularities in the vertical locations and spread of these bunching regions correspond to fluctuations in the plateau height and shape and can be attributed to finite reflectivity at both contacts[21], or to scattering of carrier waves with potential irregularities similar to that observed in a QPC[22]. A further indication of the presence of the conductance plateaus can be seen in the non-linear transport regime, around $V \sim \pm 20$ mV. At this higher bias voltage, the plateaus (bunching) at zero bias evolve into "half" plateaus when the source and drain chemical potentials cross different subbands, as reported previously in QPCs and 1D quantum wires[23,24]. From the crossing point where "half" plateaus form in Fig. 2a, we can obtain the subband spacing energy $\delta E$ to be 20~30 meV, consistent with an estimation from a 15 nm diameter cylindrical confinement potential[14].

**Josephson supercurrents**

The device was cooled in a dilution refrigerator with base temperature of 60 mK, far below the superconducting transition temperature, $T_c$, of the Al contacts ($T_c$ = 1.6 K) to allow for studies of the superconducting proximity effect. A key criterion for the observation of supercurrents in an S-N-S junction is $E_J \gg k_B T$, where $E_J$ is the Josephson junction energy[1], and $k_B$ is the Boltzman's constant. Therefore, it is important to keep the device thermally coupled to a low base temperature, as well as to reduce radiation noise fed down from the electronic circuitry to prevent electronic heating of the sample. We incorporated several stages of noise filters into the measurement system in order to keep the effective electronic temperature of the sample device close to the temperature of the mixing chamber in the dilution refrigerator (Methods and supplementary Fig. S1). The device was measured in a four-probe geometry by wire bonding a pair of wires onto each of the two Al contact pads (Fig. S1 and S2).

Fig. 2b shows a typical voltage-current ($V$-$I$) characteristic of a Ge/Si nanowire device with $L$ = 100 nm, measured in a four-probe current bias configuration. The data exhibits clear dissipationless supercurrent with zero voltage drop and effectively zero resistance as the applied current bias was swept from zero up to an $I_c$ of 113 nA (blue arrow); that is, the Ge/Si nanowire becomes superconducting with the formation of Cooper pairs inside the Ge channel due to penetration of the BCS wavefunction from the Al electrodes. Beyond $I_c$, the $V(I)$ curve abruptly switches to the normal-state dissipative conduction with a finite slope (black arrow). We note, however, that the observed $I_c$ may only represent a fraction of the actual Josephson critical current $I_{c0}$ of the junction because of premature switching caused by thermal activation[1]. When the current bias was swept down (red arrow), $V(I)$ switched back to the dissipationless state at a return current $I_r$ smaller than $I_c$. This hysteretic behavior may be the result of the phase instability due to dissipative coupling to the environment in an underdamped Josephson junction[1], or simply due to heating effect[25]. Most importantly, the $I_c$ of 113 nA and the supercurrent density of $6.4\times10^4$ A·cm$^{-2}$ are the highest observed in individual semiconductor nanowires[16], indicating the high contact quality of the Ge/Si nanowires as well as effectiveness of the noise filtering system.

**Multiple Andreev reflections**

In addition, a series of kink-like features are observed in the normal branch of the $V$-$I$ curve of Fig. 2b (green arrows). We attribute these features to multiple Andreev reflections (MAR)[5,19]. In the Andreev reflection process, an incident electron at the N-S interface becomes a Cooper pair in the condensate of the superconducting leads, resulting in a hole being coherently reflected into the normal conductor, and vice versa. In a clean conductor, this process provides the microscopic mechanism responsible for the superconducting proximity effect[5]. MARs allow for "Andreev channels" to open up in the S-N-S junction at bias voltages below the superconducting energy gap 2Δ. These Andreev channels arise from a progressive increase of the incident carrier energy as the carrier reflects between the two interfaces. Each time the electron (or hole) travels across the junction, it picks up an energy, $eV$. When the sum of the energy gain equals 2Δ of the contact leads, a resonant enhancement in the differential conductance results, which in turn exhibits as kinks in the $V$-$I$ curve[19].

Figure 3a shows a plot of $dI/dV$ as a function of the source-drain voltage bias $V$ measured using a four-probe lock-in method (Methods and Fig. S2), highlighting the kink features seen in the $V$-$I$ data. In this plot, a family of peak features were observed, symmetrically around $V$ = 0. The peak at $V$ = 0 has a height greater than $1\times10^4$ $e^2/h$, and corresponds to "infinite conductivity" when the Ge/Si NW is superconducting. The actual peak height is limited by several factors, including the applied a.c. excitation voltage, the data acquisition rate, as well as the resolution of the voltage amplifiers. The peaks at finite source-drain voltage correspond to the subharmonic energy-gap structure caused by MAR, with peak positions given by $eV_n = 2\Delta/n$ ($n$ = 1,2,3…) as expected[19]. From a fit of the MAR peak positions (Fig. 3b) we can determine that Δ = 235 μeV and find that the observed peaks correspond to $n$ = 1, 2, 3, 4, 5, 9, 13 and 25. Using the BCS relation $\Delta = 1.76 k_B T_C$ [1], the measured energy gap Δ = 235 μeV corresponds to $T_c$ = 1.6 K for the Al leads. The arrows and dashed lines in Fig. 3a mark the calculated peak positions and demonstrate the excellent fit of the observed data with theory. We note that the position of the $n$ = 1 peak at high bias does not agree with this energy gap value, likely due to heating of the junction at high bias voltages[26] that would reduce Δ. These observed subharmonic peaks are well reproducible and independent of the sweep direction. Within our measurement resolution as well as the limits of thermal broadening we did not observe any peak feature corresponding to $n$=6~8. Besides these missing peaks, it is also quite interesting to see non-consecutive high order reflections (n=9, 13, 25) with larger amplitude than the n=5 peak. An exact explanation of these observations is not yet clear. However, we note that our preliminary study of the magnetic field dependence of these subharmonic peak positions shows a systematic shift to lower energy that is consistent with the decrease in 2Δ with increasing field, and evidence that all the observed peaks originate from resonant Andreev reflections.

We are not aware of such a clear signature of MAR being reported previously. As a point of reference, the $n = 25$ subharmonic peak requires that the charge carrier traverse the channel 25 times without being back-scattered inside the channel. In addition, measurements of *dI/dV* as a function of both *V* and top gate voltage $V_g$ (Fig. 3c) show that Andreev peak positions of order $n = 2$ or higher remain constant versus $V_g$. This contrasts the complicated shifts and oscillatory behaviour of MAR peaks around a resonant level in a quantum dot[27,28], and suggests the absence of such localized states in our NWs. The higher order MARs further indicate that the contacts are quite transparent. The contact transparency estimated from BTK theory[29,30] using the *V-I* curve for $V > 2\Delta/e$ (Methods and Fig. S3) yields a value of 80 %, which is among the highest observed in superconductor-semiconductor interfaces[16,17]. This high contact transparency is maintained for different values of the gate voltage in all of the devices studied, and demonstrates that reproducible, transparent contacts can be formed using band structure engineered NW heterostructures.

In addition, we note that the device yield is quite good. We have successfully measured 18 devices with similar dimensions in a $^3$He cryostat with base temperature of 300 mK, of which 11 with normal on state resistance $R_{on}$< 20 kΩ were found to exhibit pronounced enhancement of zero bias conductance and subharmonic peak structures due to superconductivity proximity effect and MAR, while the remaining devices (with $R_{on}$ >20 kΩ) showed Coulomb blockade oscillations, presumably due to non-ideal contact interfaces. The relatively high base temperature and lack of noise filtering system in this $^3$He cryostat have precluded detailed studies of the Josephson supercurrents. However, two devices were mounted in the dilution refrigerator, and in both cases clear dissipationless supercurrents were observed, one of which is presented here. In the second device, we observed 6-8 peaks on each side of the voltage bias axis. Analysis of these data suggests missing intermediate order MAR peaks similar to the data shown above in Fig. 3(a), although because the sample was measured in a 2-probe configuration with lower resolution we do not believe it is reasonable to make an unambiguous designation of peak orders.

**Correlation between $I_c$ and normal state conductance plateaus**

The top-gated device structure allows us to vary substantially the carrier density in the Ge/Si NW heterostructures[14,15]. $V$-$I$ measurements made as a function of $V_g$ (Fig. 4a) demonstrate that $I_c$ decreases with increasing $V_g$ from -2.2 to 0.8 V; that is, $I_c$ decreases as the carrier density is reduced. These results suggest that the extent of the superconductive coupling can be tuned with the change of the carrier density inside the semiconductor NW, which thus provides an additional experimental knob that is not available in conventional S-N-S junctions employing metallic weak links[2].

We have analysed further the $V_g$ dependence of $I_c$ by comparing the measured $I_c$ and normal state conductance $G_n$ as a function of $V_g$ (Fig. 4b). The normal state conductance was obtained at zero bias while applying a 250 mT magnetic field, which completely suppresses superconductivity in the aluminium electrodes. The zero bias $G_n \equiv dI/dV$ (Fig. 4b, black curve) shows plateau features at conductance values of multiples of $2e^2/h$, similar to those observed at temperatures above $T_c$ (Fig. 2a), indicative of transport through individual 1D subbands in the Ge channel. The critical current $I_c$ (red) also shows an increase as $V_g$ is reduced, and the NW remains superconducting after $I_c$ has been "turned on" at around $V_g$ = 900 mV. This behavior contrasts previous observations in carbon nanotube S-N-S junctions where $I_c$ showed oscillatory on-off behavior due to interaction of the BCS states with resonant levels in the nanotube[27]. Significantly, our measured $I_c$ increases step-wise with a step height of $\delta I_c \sim$ 20 nA, strongly suggestive of quantization in superconducting critical current in our Ge/Si NW heterostructure.

The correlation between the plateau structures at quantized values in the $I_c$-$V_g$ and $G_n$-$V_g$ curves can in fact be confirmed by plotting the ratio of $I_c$ and $G_n$, or the $I_c R_n$ product vs. $V_g$ as shown in Fig. 4c, where $R_n \equiv 1/G_n$ is the normal state resistance. The $I_c R_n$ product is essentially a constant at around 200 µV, over a wide range of gate voltages, indicating that the steps in $I_c$ occur when individual subbands are populated in the NW. We further note that the ratio of the measured $I_c R_n$ with $\Delta/e$ reaches 85 % and is substantially higher than previous studies using InAs NWs and carbon nanotubes, which yield $I_c R_n/(\Delta/e)$ ratios of only 15 to 47%[16,27].

These results can be compared with theoretical predictions. In the clean limit (mean free path $l > \xi_0$), the Ge/Si nanowire coherence length is $\xi_0 = \hbar v_F / \pi \Delta_0$ = 140 nm, where the Fermi velocity $v_F$ = 1.5 x 10$^5$ m/s is estimated by assuming a Fermi energy of 20 meV on the first subband, and $\xi_0$ will increase for higher energy subbands. The Ge/Si NW 1D channels in our measurements thus fall into the regime $L << \xi_0$ where theory[4] predicts that $I_c R_n = \pi \Delta/e$ and that the critical current increases in steps of $\delta I_c = e\Delta/\hbar$. The measured $I_c R_n$ = 200 µV differs from $\pi \Delta/e$ = 738 µV by a factor of ~ 3.6. This discrepancy can be explained by premature switching due to thermal activation in a capacitively and resistively shunted junction[1], which would lead to a measured $I_c$ smaller than the actual Josephson critical current $I_{c0}$. Consistent with this explanation we also observe a value of $\delta I_c$ = 20 nA that is a factor of 3 lower than the predicted value $\delta I_c = e\Delta/\hbar$ = 57 nA. Note that the finite contact transparency, as well as the limited $L/\xi_0$ value may also contribute to the observed reduction factor in $\delta I_c$ and $I_c R_n$. We expect that optimization of the junction geometry by, for example, using a shorter junction, minimizing the stray capacitance from the contact leads and improving the measurement to further reduce the sample effective temperature, will yield values of $I_c$ closer to the universal predictions. Despite these limitations, the work presented here is the first study of the superconducting proximity effect coupled with discrete 1D modes in the $L \ll \xi_0$ regime, while previous efforts in QPCs with $L \gg \xi_0$ achieved $\delta I_c$ values two orders of magnitude smaller (vs. a factor of 3-4 in our work) than the universal value of $e\Delta/\hbar$ [6,7].

In summary, we have presented studies of mesoscopic Josephson junctions using Ge/Si NW heterostructures that demonstrate tunable dissipationless supercurrents as well as clear signatures of multiple Andreev reflections. Significantly, by tuning the carrier density to populate discrete 1D modes we show a distinct correlation between the quantized normal state conductance and steps of the critical current. The smooth, scattering-free 1D confinement potential in the Ge/Si nanowire channel, along with the local top gate device structure, will allow the use of Josephson junctions as tunable superconducting field-effect transistors and unique quantum interference devices. More generally, the ability to couple low-dimensional systems with superconducting phenomena opens new opportunities in this interesting mesoscopic regime.

**Methods**

**Fabrication of Ge/Si nanowire devices**

The epitaxial core/shell Ge/Si NWs with an average core diameter of 14.6 nm and Si shell thickness of 1.7 nm and <110> core growth direction used to fabricate devices were prepared as described previously[14,15,20]. Al source (S)/drain (D) contact electrodes (30 nm thickness) were defined by electron beam lithography and deposited by thermal evaporation. The nanowires and S/D electrodes were then covered with a 10 nm $HfO_2$ high dielectric constant layer (80 cycles using atomic layer deposition (ALD))[15], and then electron beam lithography was used to define the top gate, followed by thermal evaporation of Cr/Au (5/50 nm).

**Low-noise measurement setup**

We employed three stages of filtering on all the electrical leads. An illustration of the experimental setup is shown in Supplementary Fig. S1. At room temperature, π filters (Spectrum Control) are placed on each electrical lead before they enter the cryostat. The π filters are mounted inside a Pomona box (Pomona Electronics) through a RF-tight copper divider. The copper divider separates the 'in' section from the 'out' section to further shield the outgoing signal from radiation. Above 1 MHz, the measured attenuation of these π filters is 80 dB. A second stage of filtering is provided by 10 kΩ metal film resistors embedded in a copper chuck. The copper chuck is thermally anchored to the mixing chamber. The 10 kΩ metal film resistors, along with the cable inductance and capacitance, serve as cold low-pass RLC filters with an estimated noise cut off frequency of 10 kHz. The final stage of filtering of the leads, effective for frequencies above 1 GHz, is provided by a copper powder filter that is integrated on the cold finger of the dilution refrigerator. The area between the inner and outer tubes of the cold finger is filled with a mixture of surface-oxidized copper powder and epoxy (Stycast #1266, Emerson and Cumming), and the electrical leads were coiled inside this cavity before being connected to the sample. Finally, the $^3$He/$^4$He dilution refrigerator and the measurement electronics sit inside a sealed shielded room. The data acquisition computer is placed outside the shielded room, and communicates with the measurement electronics via optical fibers.

Four-probe measurements, as depicted in Fig. S2, were used to obtain voltage-bias measurements of $dI/dV$. The four probes were formed by wire bonding a pair of wires onto each aluminum contact pad.

**Estimation of contact transparency using the BTK theory**

At voltage larger than the BCS gap $2\Delta$, transport can occur via the excited carriers in the leads and is no longer superconducting in nature. This can be seen in Fig. S3, where a fit to the $V$-$I$ curve at the normal state above $2\Delta$ extrapolate to a finite excess current $I_{exc}$ (red solid line) and does not go to the origin. This is a clear manifestation of the conductance enhancement below $V = 2\Delta$ due to superconducting proximity effect. The slope of the line yields the normal state resistance $R_n$. We analyzed $I_{exc}$ and $R_n$ under the framework of the standard BTK theory[29,30]. In this particular case, $I_{exc}$ = 78 nA, $R_n$ = 3.5 kΩ. Using $\Delta$ = 235 μeV we obtain $e \cdot I_{exc} \cdot R_n/\Delta$ = 1.16, which corresponds to a contact transparency of 80%. Similar values for the contact transparency are obtained at different gate voltages. We note that this calculation of contact transparency is a conservative estimate as it does not include the reduction of BCS gap at high bias, as manifested by the shift of $n$=1 MAR peak. For example if the gap energy is reduced to 85% of the value at lowest temperature, in our case the $e \cdot I_{exc} \cdot R_n/\Delta$ would increase from 1.16 to 1.36, and the obtained contact transparency would increase to 85% from 80%.

**Author contributions** J.X. and A.V. performed the experiments and analyzed the data with help from M.T., all authors discussed the results and co-wrote the manuscript.

**Acknowledgements** We thank Y.J. Doh and J.U. Free for helpful discussions. M.T. acknowledges support from the National Science Foundation. R.M.W. acknowledges support of this work by DARPA-QuIST, and the Nanoscale Science and Engineering Center at Harvard University. C.M.L. acknowledges support of this work by the Defense Advanced Research Projects Agency, Army Research Organization and NSF.

Correspondence and requests for materials should be addressed to R.M.W. or C.M.L.

**Competing financial interests** The authors declare that they have no competing financial interests.

**Figure Legends**

**Figure 1.** Schematic diagram and device image of a Ge/Si nanowire device. **a**, Schematic of a Ge/Si core/shell nanowire heterostructure. Also shown is the radial wave function Ψ(r) for the lowest 1D subband in a cylindrical quantum wire, where $\lambda_F$ is comparable to the channel diameter *w*. **b**, Cross-sectional diagram showing the hole gas in Ge quantum well confined by the epitaxial Si shell, where VB is the valence band. **c**, Scanning electron microscopy image of a representative top-gated device. The dashed line highlights the position of the nanowire. **d**, Cross-sectional schematic of the device along the dashed line in **c** showing the Al source(S)/drain(D) contact electrodes, $HfO_2$ dielectric and Cr/Au top gate.

**Figure 2.** Transport characteristic of Ge/Si nanowire above and below $T_c$. **a**, *dI/dV-V* plots recorded at $V_g$ = 0.8 to -3.5V from bottom to up in 50 mV steps with no offset applied (*T* = 10 K). An offset resistance of 5 kΩ was subtracted from data to account for the fixed series resistance in this 2-probe measurement as well as effects arising from the finite contact resistance and scattering with impurity potentials inside the nanowire. Coulomb charging effects were not observed at temperatures down to at least 5 K for all gate voltages. **b**, Four-probe *V-I* data measured for a Ge/Si nanowire at $V_g$= -3.5 V at the dilution refrigerator base temperature of 60 mK. Black and red curves correspond to different current sweeping directions as indicated by arrows. The horizontal green arrows point to the kink structures due to multiple Andreev reflections. Blue vertical arrows point to the positions of $I_c$ and $I_r$.

**Figure 3.** Multiple Andreev reflection in the Ge/Si nanowire device. **a**, The *dI/dV* vs. *V* data were taken in a four-probe geometry (see Supplementary Fig. S2) at $V_g$= -900 mV. Arrows and dashed lines mark the theoretical MAR peak positions at $V_n=2\Delta/n \cdot e$ with values of *n* listed. **b**, Plot of the MAR peak position vs. the inverse index 1/*n*. Red line is a linear fit of all the points (except $n = 1$) through the origin and the slope is proportional to the gap energy. **c**, $dI/dV$-$V$-$V_g$ plot showing the MAR features at different gate voltages. The arrows and numbers at the top indicate the corresponding $n^{th}$ order MAR features.

**Figure 4.** Gate voltage dependence of critical current and normal conductance. **a**, $V$-$I$ curves for the same device measured at different gate voltages. All curves were taken while sweeping current from zero toward 100 nA. **b**, Normal state $dI/dV$ vs. $V_g$ curve (black) and $I_c$ vs. $V_g$ (red). The normal state conductance was measured via four-probe geometry at $V_{bias} = 0$ with applied magnetic field $H = 250$ mT, a series resistance of $R_s = 200$ Ω has been subtracted. $I_c$ is extracted from individual $V$-$I$ measurement as the onset of voltage switching in the direction as the black arrow in Fig. 2b. **c**, Product of $I_c$ and $R_n$ vs. gate voltage, where $R_n$ is extracted from the slope of the individual $V$-$I$ curve at voltages larger than $2\Delta/e$, or 470 μV.

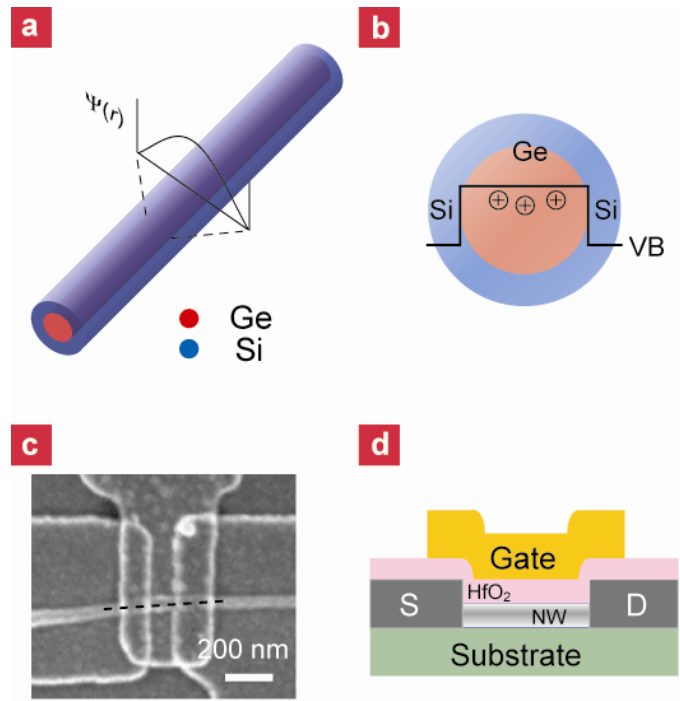

Figure 1

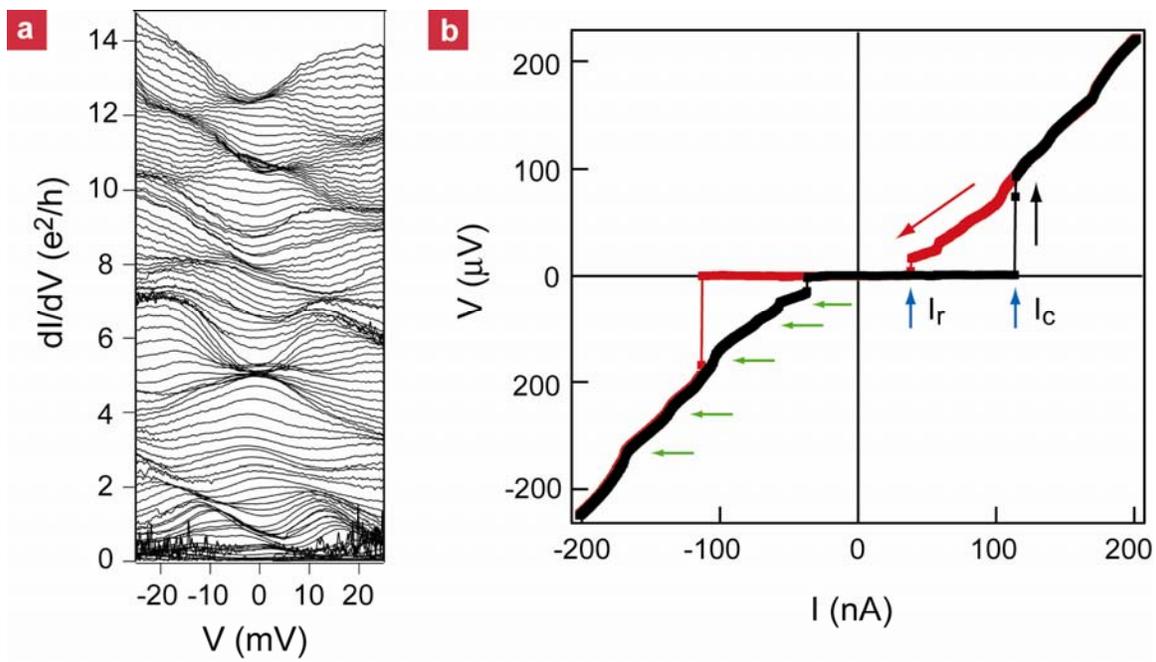

**Figure 2**

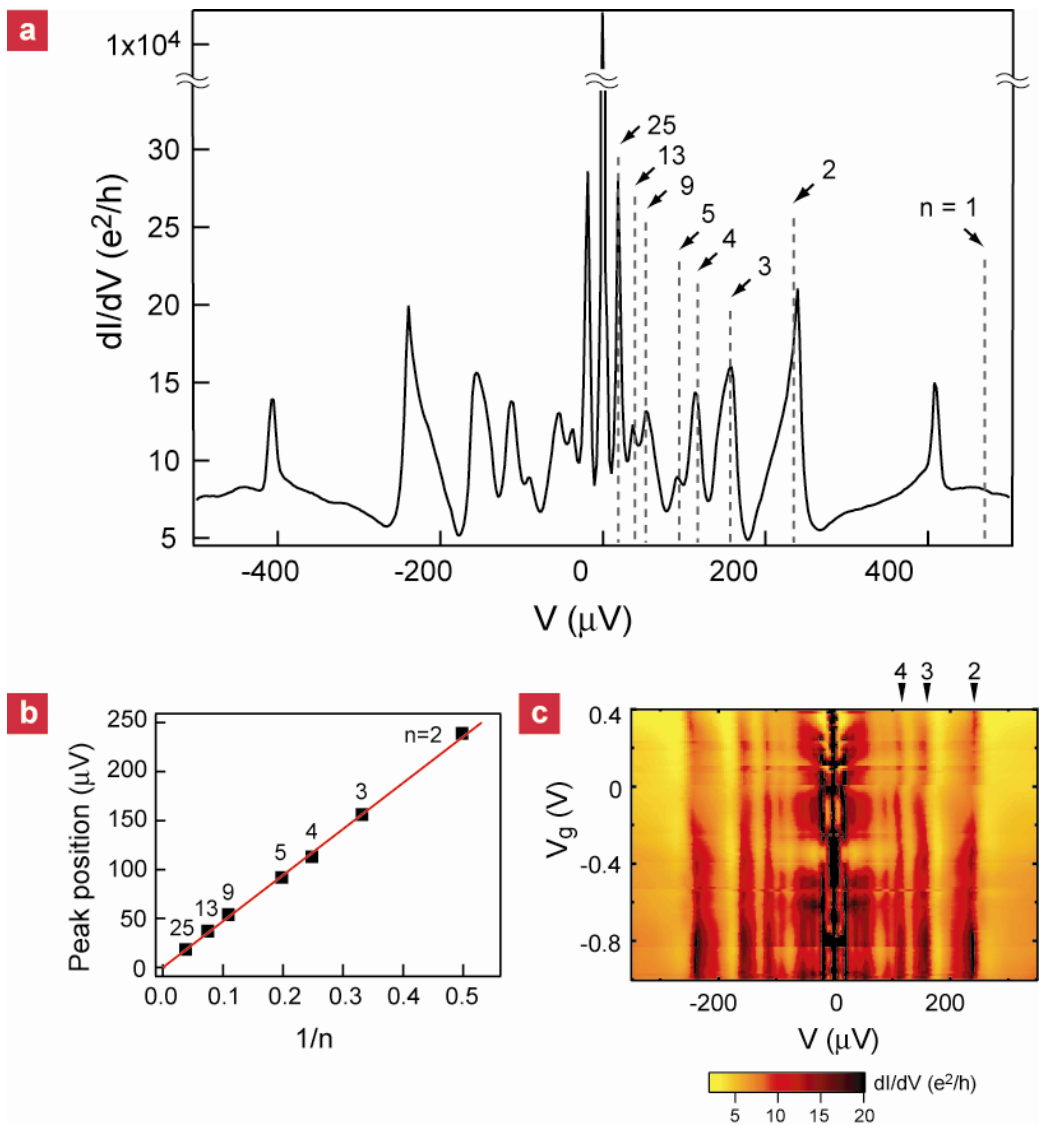

**Figure 3**

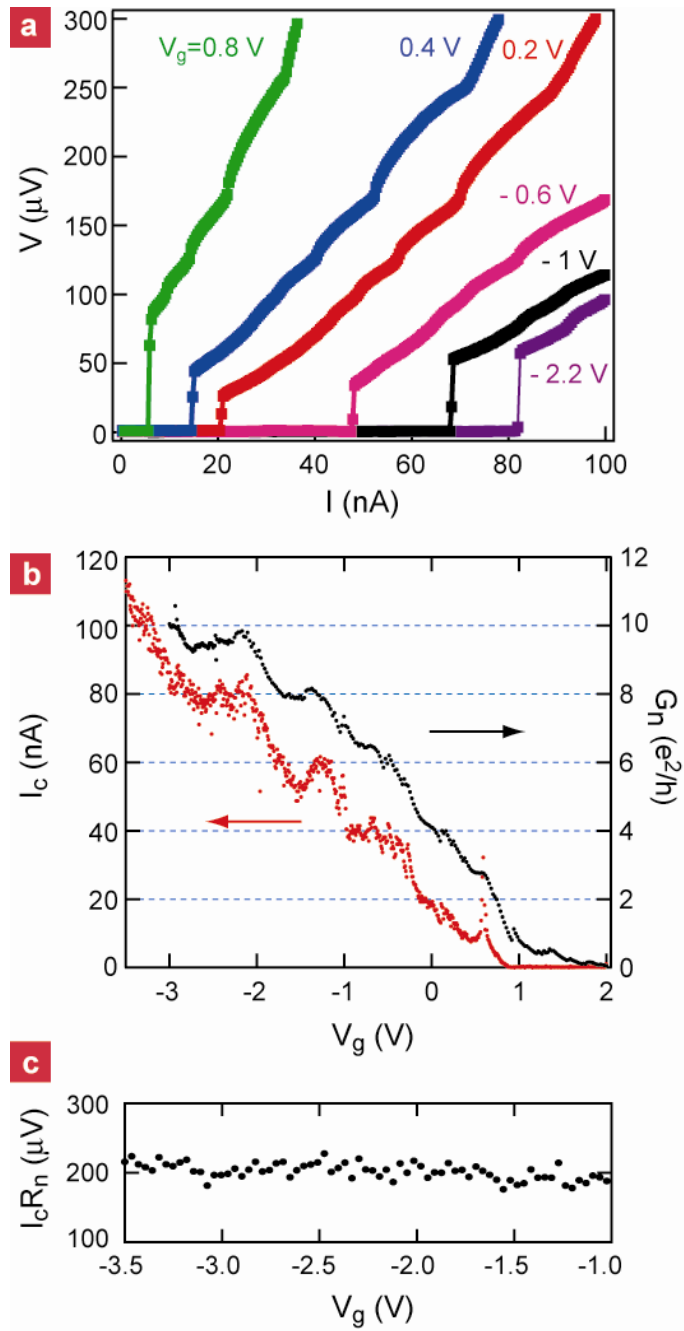

**Figure 4**

# Supplementary Information

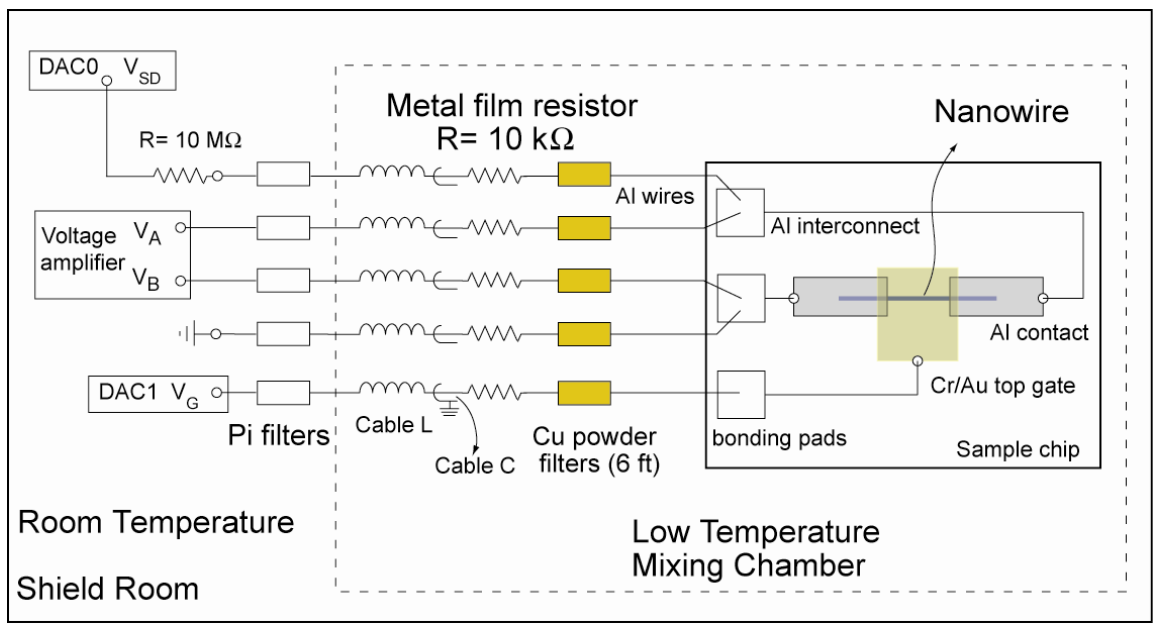

**Figure S1.** Schematic diagram of a typical current bias nanowire measurement showing the three stages of filtering. Each electrical lead passes through a $\pi$ filter placed on the top of the dilution insert. Metal film resistors and copper powder filters thermally anchored to the mixing chamber provide additional noise attenuation. The dilution refrigerator and measurement electronics are placed inside a shielded room.

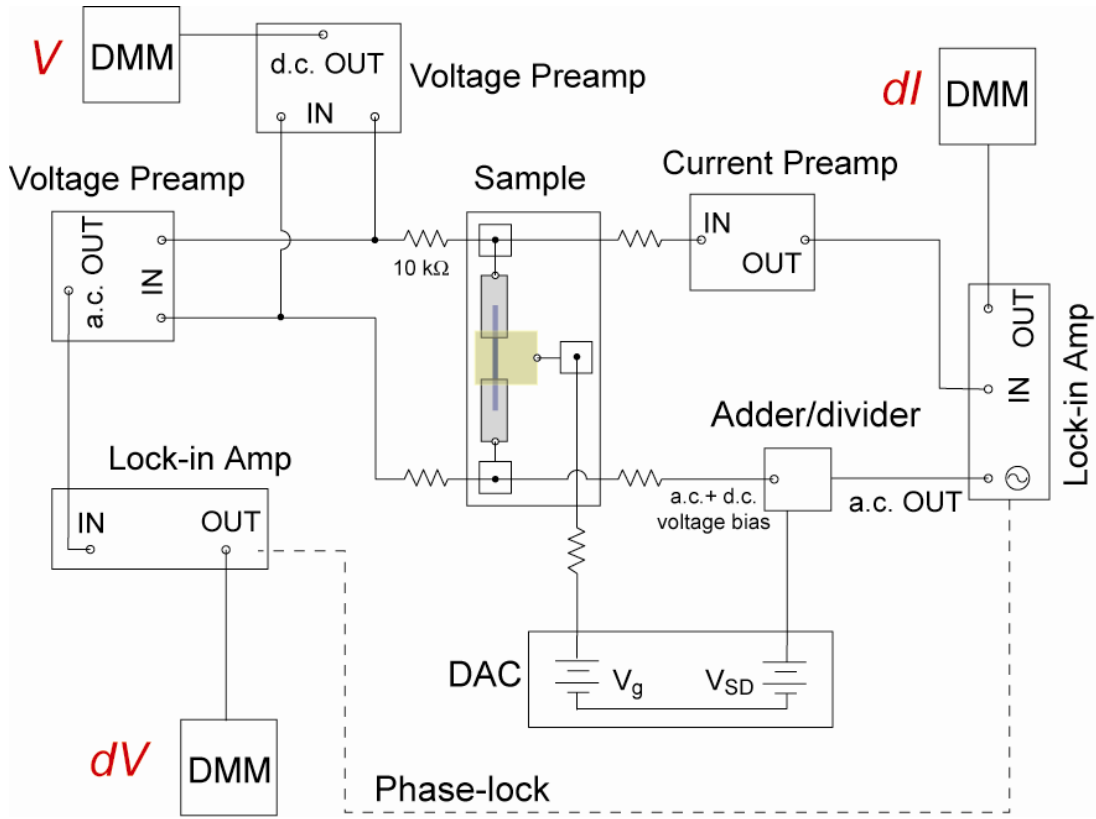

**Figure S2.** Circuit diagram showing the 4-probe voltage bias nanowire measurement setup. During measurement each *dI/dV* vs. *V* data point consists of reading *V*, *dI* and *dV* separately from the three digital multimeters (DMMs),

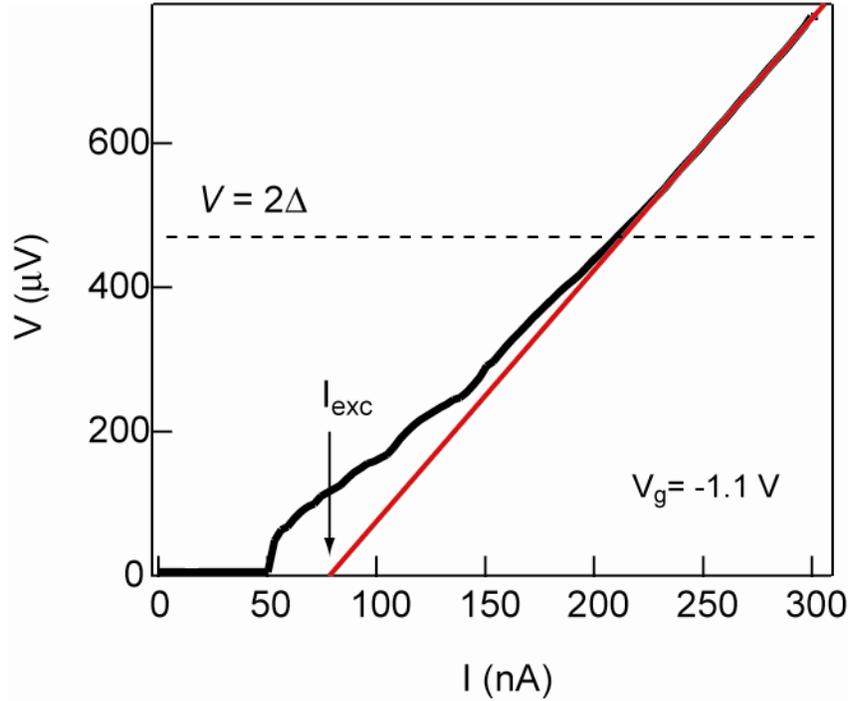

**Figure S3.** *V-I* curve (black line) measured at $V_g$ = -1 V. *V(I)* returns to the normal state only when $eV > 2\Delta$ (black dashed line). The value of $2\Delta$ is determined from multiple Andreev reflection data and equals 470 µeV. In the high bias regime, a straight line is fitted to *V(I)* (red solid line) and the normal state resistance $R_n$ and excess current $I_{exc}$ can be determined from the slope and intercept, respectively. The excess current allows for an approximate value of the contact transparency to be extracted (see text).